\documentclass[11pt]{article}

\input epsf
\usepackage{latexsym}
\usepackage{amssymb}
\usepackage[dvips]{graphicx}

\input epsf
\usepackage{latexsym}
\usepackage{amssymb}
\usepackage{amsmath}
\usepackage[dvips]{graphicx}

\begin{document}
\centerline{\Large \bf Perturbations in stochastic inflation}

\vskip 2 cm

\centerline{Kerstin E. Kunze
\footnote{E-mail: kkunze@usal.es, Kerstin.Kunze@cern.ch} }

\vskip 0.3cm

\centerline{{\sl Departamento de F\'\i sica Fundamental,}}
\centerline{and}
\centerline{{\sl Instituto Universitario de F\'\i sica Fundamental y Matem\'aticas
(IUFFyM),}}
\centerline{{\sl Universidad de Salamanca,}}
\centerline{{\sl Plaza de la Merced s/n, E-37008 Salamanca, Spain }}

\vskip 1.5cm

\centerline{\bf Abstract}
\vskip 0.5cm
\noindent
The perturbative approach to stochastic inflation 
is used to determine the spectrum of density fluctuations 
and gravitational waves due to the coarse grained field.
The amplitude of the curvature fluctuation spectrum, the spectral index 
and the running of the spectral index are in general found to
be smaller than in the standard approach to inflation.
Furthermore, the amount of non-gaussianity
due to the second order perturbation 
is estimated.

\vskip 1cm

\section{Introduction}

Inflation provides a mechanism in order 
to seed large scale structure formation.
Quantum fluctuations of the inflaton are stretched beyond
the horizon and are converted into classical perturbations.
In the standard approach to calculating the 
spectrum of the resulting fluctuations expectation 
values of the quantum fluctuations 
are identified with statistical averages 
of classical perturbations.
Thus the amplitude of fluctuations is determined by the
vacuum expectation value of the quantum field.
The details of the quantum to classical transition
are neglected.
It is assumed that perturbations are classical
instantaneously at horizon exit.
The assumption of a classical behaviour outside the 
horizon has been found to be justified, as it is 
related to the particular behaviour of the 
mode functions in a de Sitter space-time (see for example
\cite{ll}).

In the stochastic approach to inflation the dynamics
of the quantum to classical transition is 
effectively described by a classical noise.
The scalar field is coarse grained. Thus the field
modes are split into a subhorizon part and a 
superhorizon part. The superhorizon part constitutes
the classical, homogeneous coarse grained field driven 
by the stochastic noise due to the subhorizon 
modes leaving the horizon.
As the inflationary universe keeps expanding
rapidly more and more short wavelength modes
are stretched beyond the horizon and subsequently
contribute to the coarse grained, classical
field. 
The coarse graining scale is at least of the order of the 
horizon. 
As shown in \cite{sto1} in the case of a
scalar field in a de Sitter space-time, the 
effective dynamics of the classical, coarse grained
field is described by a Langevin equation with 
a white noise term.
However, it has been shown that the appearance 
of the white noise term is a direct consequence
of the choice of window function to 
separate the sub- and superhorizon modes \cite{wv}.
Furthermore, models of stochastic inflation 
attempting to model in more detail the decoherence
of the quantum field can be found, for example, 
in \cite{more-mod}.

Here, however, only the simplest model of stochastic
inflation will be considered \cite{sto1}.
Thus the dynamics of the coarse grained field is 
determined by a white noise term.
Recently, the equations of stochastic inflation
for the coarse grained inflaton
have been solved explicitly in a perturbative way
\cite{sto-pert} (for earlier work in this context, 
see \cite{glmm}). This allows to 
take into account as back reaction the
evolution of the inflaton, namely the time variation of the 
Hubble parameter.
In this work the perturbative approach is used to
determine the spectrum of the resulting classical
fluctuations due to the coarse grained field. 
Moreover, it is possible to estimate the amount
of non-gaussianity, encoded in the parameter 
$f_{NL}$ \cite{fnl}-\cite{sl}, by including terms 
up to second order in the perturbative expansion.

\section{Curvature perturbation}

In the slow roll regime of stochastic inflation
the evolution of the coarse grained scalar 
field $\varphi$ is described by the Langevin equation
\cite{sto1},
\begin{eqnarray}
\dot{\varphi}+\frac{1}{3H}\frac{dV}{d\varphi}=\frac{H^{\frac{3}{2}}}
{2\pi}\xi(t),
\label{e1}
\end{eqnarray}
where $\xi(t)$ is a white noise process with
$\langle\xi(t)\rangle=0$ and $\langle\xi(t)\xi(t')\rangle=\delta(t-t')$.
Furthermore, the Hubble parameter is given by
$H^2(\varphi)=\frac{8\pi}{3M_P^2}V(\varphi)$.
Following \cite{sto-pert} equation (\ref{e1})
can be solved by the perturbative expansion 
\begin{eqnarray}
\varphi(t)=\varphi_c(t)+\delta\varphi_1(t)+\delta\varphi_2(t)+...,
\label{e2}
\end{eqnarray}
where $\varphi_c$ is the deterministic background value
of the coarse grained field.
Using the expansion (\ref{e2}) in equation (\ref{e1})
results in \cite{sto-pert}
\begin{eqnarray}
\frac{d\varphi_c}{dt}&=&-\frac{M_P^2}{4\pi}H'(\varphi_c)
\label{pp1}\\
\frac{d\delta\varphi_1}{dt}&=&-\frac{M_P^2}{4\pi}H''(\varphi_c)
\delta\varphi_1+\frac{H^{\frac{3}{2}}(\varphi_c)}{2\pi}
\xi(t)
\label{pp2}\\
\frac{d\delta\varphi_2}{dt}&=&-\frac{M_P^2}{4\pi}H''(\varphi_c)
\delta\varphi_2-\frac{M_P^2}{8\pi}H'''(\varphi_c)\left(
\delta\varphi_1\right)^2
+\frac{3}{4\pi}H^{\frac{1}{2}}(\varphi_c)
H'(\varphi_c)\delta\varphi_1\xi(t),
\end{eqnarray}
where a prime denotes the derivative with respect to 
$\varphi_c$.
The solutions for the first and second order perturbations 
are given by \cite{sto-pert}
\begin{eqnarray}
\delta\varphi_1(t)&=&\frac{H'[\varphi_c(t)]}{2\pi}
\int_0^td\tau\frac{H^{\frac{3}{2}}[\varphi_c(\tau)]}{H'[\varphi_c(\tau)]}
\xi(\tau)
\label{p1}
\\
\delta\varphi_2(t)&=&H'[\varphi_c(t)]
\int_0^t d\tau\left[-\frac{M_P^2}{8\pi}\frac{H'''[\varphi_c(\tau)]}
{H'[\varphi_c(\tau)]}\left(\delta\varphi_1\right)^2
+\frac{3}{4\pi}H^{\frac{1}{2}}[\varphi_c(\tau)]\delta\varphi_1\xi(\tau)
\right].
\label{p2}
\end{eqnarray}

The spectrum induced by the first order fluctuation
$\delta\varphi_1$ can be calculated using the following
expression for the curvature fluctuation, e.g. \cite{ll},
\begin{eqnarray}
\langle{\cal R}^2\rangle=\left(
\frac{H(\varphi_c)}{\dot{\varphi_c}}\right)^2
\langle\delta\varphi_1^2\rangle.
\end{eqnarray}
Besides, since the curvature perturbation is constant 
outside the horizon it can be assumed that its spectrum is
just a function of wave number.
Thus
\begin{eqnarray}
\langle{\cal R}^2\rangle=\int_0^{k}{\cal P}_{\cal R}
\frac{dq}{q}=\int_{-\infty}^{\ln k}{\cal P}_{\cal R}
d\ln q. 
\end{eqnarray}
where $k$ is some maximal scale, which corresponds to a minimal
length scale that can be ``probed''. This scale is determined 
by the coarse graining scale. 
The spectrum is calculated at the coarse graining scale,
$k=\sigma aH$, where $\sigma$ is a parameter. 
Thus with $d\ln k=Hdt$ the spectrum is given by 
\begin{eqnarray}
{\cal P}_{\cal R}=\frac{1}{H}\frac{d}{dt}\langle{\cal R}^2\rangle,
\label{spec}
\end{eqnarray}
see for example \cite{mat}.
Using the solution for $\delta\varphi_1$ results in 
\begin{eqnarray}
{\cal P}_{\cal R}=
\left(\frac{4\pi}{M_P^2}\right)^2
\left(\frac{H}{H'}\right)^2
\frac{H^2}{4\pi^2}
-\frac{8}{M_P^4}(H')^2\int_{\varphi_c}^{\varphi_0}
d\psi_c\left(\frac{H}{H'}\right)^3,
\label{pP1}
\end{eqnarray}
where $\varphi_0\equiv\varphi_c(0)$.
Clearly, the first term is the standard result of
the curvature perturbation spectrum due to a quantum field in a 
de Sitter background.
The correction term is due to the evolution of
the Hubble  parameter in time. As can be shown, assuming
$H$ to be constant in time results in the correction
term vanishing.
In the standard approach the change due to the 
evolution of the inflaton is neglected. The variation in 
$H$ is being considered only when deriving the
spectral index and derived quantities of that (see
for example \cite{ll}).
Moreover, choosing $\varphi_0$ to be the value of the 
coarse grained field at horizon crossing, it is found that 
for the choice of the coarse graining scale to be of the 
order of the horizon there is no correction of the 
spectral amplitude. This is consistent with the 
picture that the spectrum in the standard 
approach is calculated at horizon crossing.
In case that the coarse graining scale and the horizon 
are not the same equation (\ref{pP1}) implies that the
spectrum is reduced with respect to the standard result.
Modes with wavelengths between the horizon and the 
coarse graining scale are not taken into account. Thus 
the amplitude of the spectrum in the stochastic approach 
is in general reduced with respect to the amplitude in 
the standard approach.

In the following the potential is assumed to be of the 
form 
\begin{eqnarray}
V(\varphi_c)=V_0\left(\frac{\varphi_c}
{M_P}\right)^n.
\label{pot}
\end{eqnarray} 
The mean square fluctuation at first order is found to
be 
\begin{eqnarray}
\langle\delta\varphi_1(t)^2\rangle=\frac{H^2}{4\pi^2}
\frac{2\pi}{n}\left(\frac{\varphi_c}{M_P}\right)^2
\left[\left(\frac{\varphi_0}{\varphi_c}\right)^4-1\right].
\end{eqnarray}
In comparison, the amplitude of quantum fluctuations 
of a massless free scalar field in 
de Sitter space-time is given by $\delta\phi=H/2\pi$.
However, it can be shown that at a time $t=H^{-1}$
the mean square first order fluctuation is given by
$\langle\delta\varphi_1^2(H^{-1})\rangle\simeq\frac{H^2}{4\pi^2}$.
This is consistent with the general picture that the 
stochastic differential equation (\ref{e1})
describes a random walk process with a step size of
the order of $H/2\pi$ over a characteristic time 
$H^{-1}$ \cite{et-inf}. Here, this holds at first 
order in the expansion of the perturbation.

The power spectrum of the curvature fluctuations ${\cal P}_{\cal R}$
for the potential (\ref{pot}) is found to be 
\begin{eqnarray}
{\cal P}_{\cal R}=\frac{32\pi}{3n^2}\left(\frac{V_0}{M_P^4}\right)
\left(4-n\left[\left(\frac{\varphi_0}{\varphi_c}\right)^4-1\right]
\right)\left(\frac{\varphi_c}{M_P}\right)^{n+2}.
\label{n1}
\end{eqnarray}
This can also be written as
${\cal P}_{\cal R}=\frac{2}{3}\left(4
-n\left[\left(\frac{\varphi_0}{\varphi_c}\right)^4-1\right]
\right)\frac{1}{\epsilon}\left(\frac{V}{M_P^4}\right)$,
where $\epsilon\equiv\frac{M_P^2}{16\pi}\left(\frac{V'}{V}
\right)^2$ is one of the slow roll parameters.
In this case the spectrum depends explicitly on the 
value $\varphi_0$ of the coarse grained field.
Besides, the requirement ${\cal P}_{\cal R}>0$ imposes 
an upper bound on the fraction $\left(\frac{\varphi_0}{\varphi_c}
\right)$, implying the range
$1\leq\left(\frac{\varphi_0}{\varphi_c}
\right)<\left(1+\frac{4}{n}\right)^{\frac{1}{4}}$.
Thus this gives an upper bound on the size of the 
coarse graining domain. However, in principle, there should not
be any limit. The upper cut-off is due to the fact that this 
is a perturbative approach and here only the first order is
considered.
Moreover, considering times of the order of the 
Hubble time leads to a change in the coarse grained
scalar field of the order
\begin{eqnarray}
\frac{\varphi_0}{\varphi_c}\simeq 1+\frac{n}{8\pi}
\left(\frac{\varphi_c}{M_P}\right)^{-2},
\end{eqnarray}
which is much smaller than the maximal bound imposed
by the amplitude of the spectrum.
The stochastic equation (\ref{e1}) is derived for 
de Sitter space-time. Using the stochastic approach to
inflation to find the effect on the 
spectrum due to the evolution of the Hubble parameter
can be interpreted as a perturbation around the
de Sitter case. This is done varying the
size of the coarse graining domain, that is
varying the ratio $\frac{\varphi_0}{\varphi_c}$.

\begin{figure}[ht]
\centerline{\epsfxsize=2.2in\epsfbox{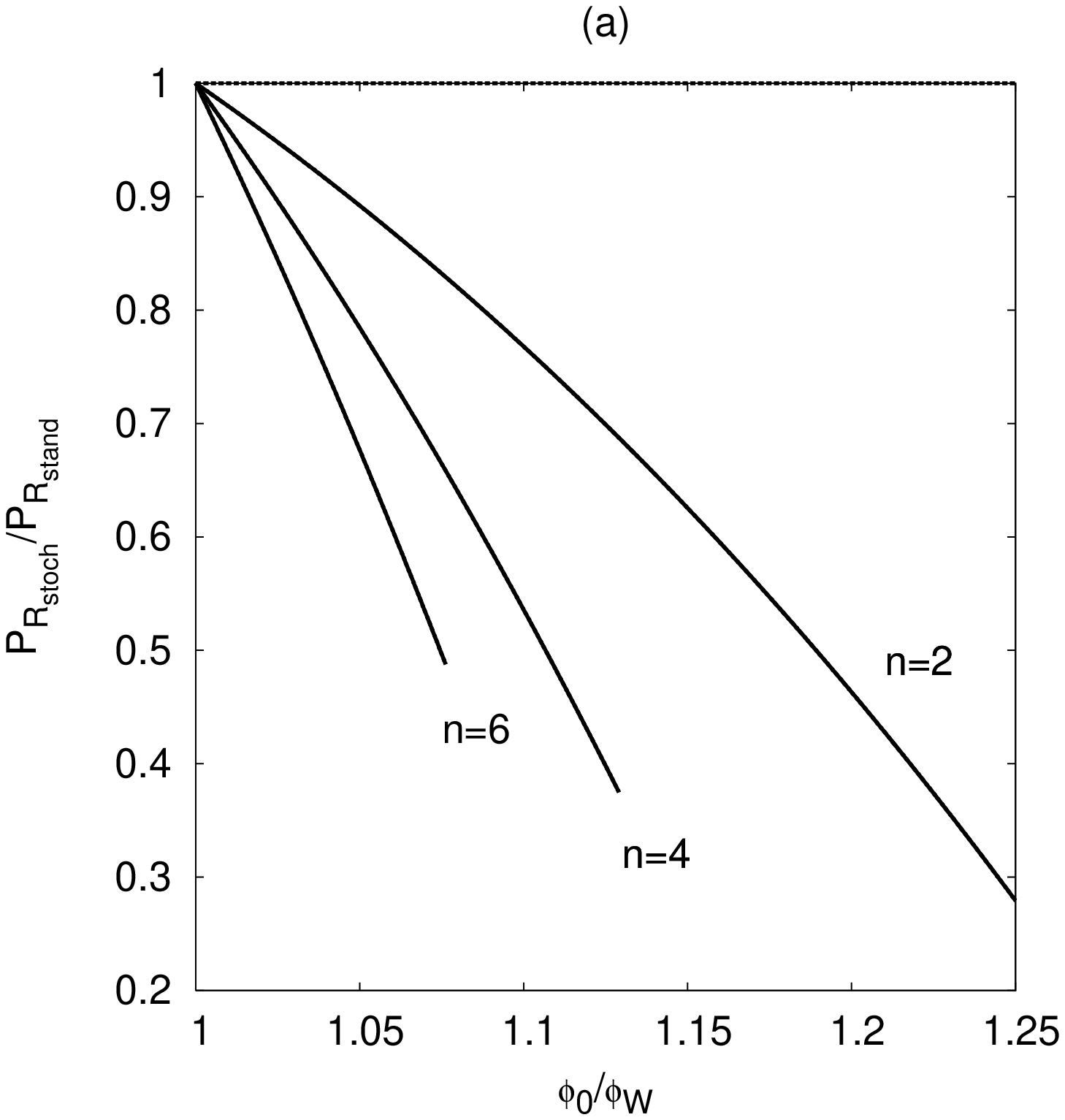}
\hspace{0.05cm}
\epsfxsize=2.2in\epsfbox{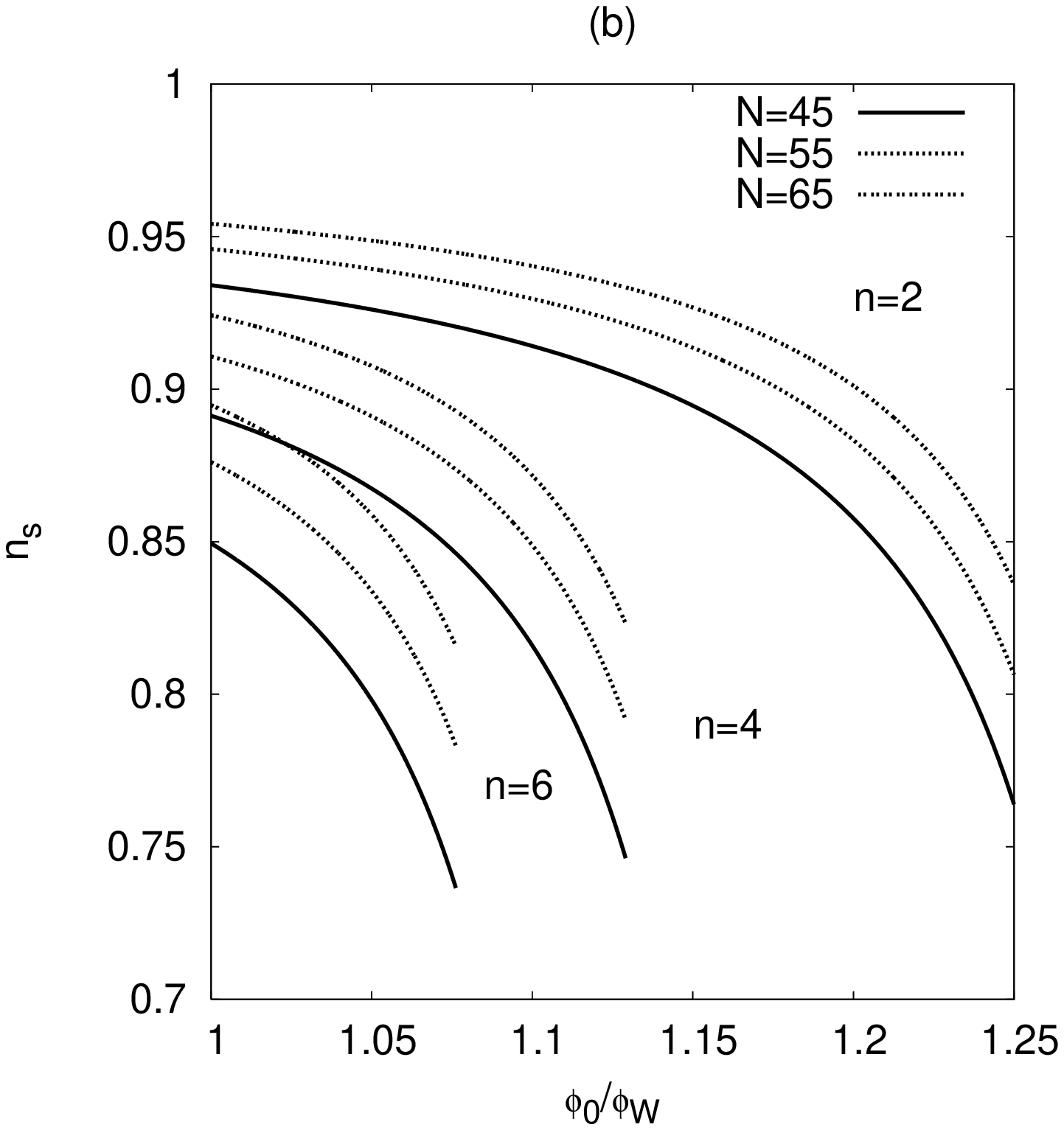}
\hspace{0.05cm}
\epsfxsize=2.2in\epsfbox{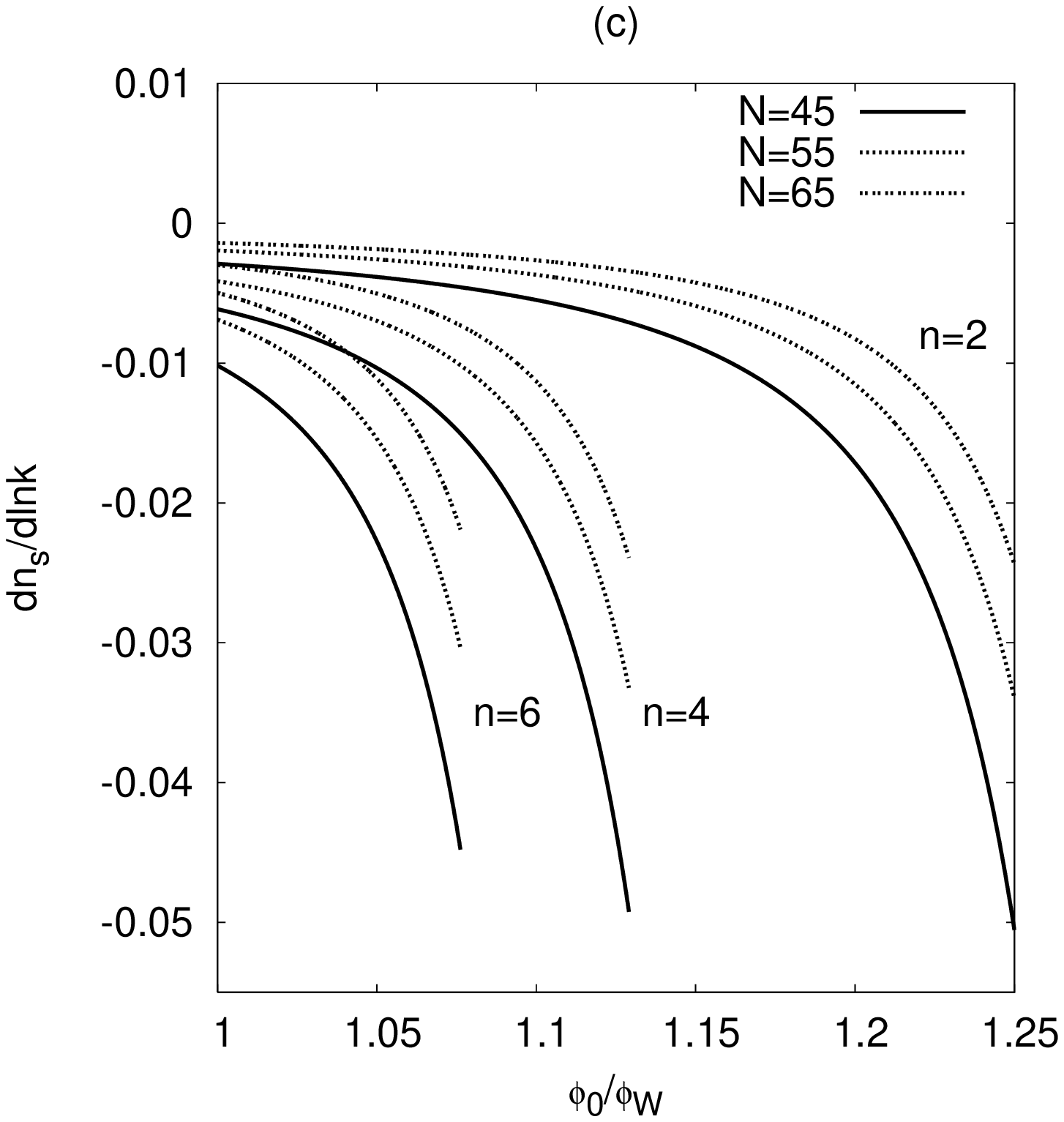}
}
\caption{(a) The ratio of the amplitude 
of the curvature perturbation due to the 
stochastic over the amplitude due to the standard approach
$\frac{{\cal P}_{{\cal R}_{stoch}}}{{\cal P}_{{\cal R}_{stand}}}$.
(b) The spectral index in the stochastic approach.
(c) The running of the spectral index in the stochastic approach.
Graphs are shown for $N_W=45,55,65$. Furthermore, $n$ denotes the
power of the potential.
}
\label{fig1}
\end{figure}

For comparison, the spectrum in the standard approach is given by 
$
{\cal P}_{\cal R}=\frac{8}{3M_P^4}\frac{V}{\epsilon}$ \cite{ll}.
Expression (\ref{n1}) reduces to this for
$\left(\frac{\varphi_0}{\varphi_c}\right)=1$.
The ratio of the amplitude of the curvature fluctuations 
in the stochastic approach over the one in the standard approach 
is shown in figure \ref{fig1} (a).
For $\left(\frac{\varphi_0}{\varphi_c}\right)>1$,
the amplitude in the stochastic approach is always lower than in 
the standard approach.

Whereas the spectral index $n_s$ given by
\begin{eqnarray}
n_s=1+\frac{d\ln{\cal P}_{\cal R}}{d\ln k}
\end{eqnarray}
is found to be
\begin{eqnarray}
n_s=1-\frac{n}{8\pi}
\frac{(n+2)(n+4)-n(n-2)\left(\frac{\varphi_0}{\varphi_c}\right)^4}
{n+4-n\left(\frac{\varphi_0}{\varphi_c}\right)^4}
\left(\frac{\varphi_c}{M_P}\right)^{-2},
\label{n2}
\end{eqnarray}
the running of the spectral index is given by
\begin{eqnarray}
\frac{dn_s}{d\ln k}=-(n_s-1)^2+
\frac{n^2}{64\pi^2}
\frac{n(n+2)(n+4)-n(n-2)(n-4)\left(\frac{\varphi_0}{\varphi_c}
\right)^4}
{n+4-n\left(\frac{\varphi_0}{\varphi_c}\right)^4}
\left(\frac{\varphi_c}{M_P}\right)^{-4}.
\label{n3}
\end{eqnarray}
The quantities (\ref{n1}) to (\ref{n3}) are calculated 
at $\varphi_c=\varphi_W$ which is the field value
$N_W$ e-folds before the end of inflation, when the 
characteristic scale of WMAP is of the order of the size of the 
coarse graining domain.
$\varphi_W$ is given by
\begin{eqnarray}
\left(\frac{\varphi_W}{M_P}\right)=\sqrt{\frac{n}{4\pi}
N_W+\frac{n^2}{16\pi}}.
\label{pw}
\end{eqnarray}
The spectral index in the standard approach \cite{ll},
$n_s=1-6\epsilon+2\eta$, with
$\eta\equiv \frac{M_P^2}{8\pi}\frac{V''}{V}$ the 
second slow roll parameter, is given by
\begin{eqnarray}
n_{s,stand}=1-\frac{n}{8\pi}(n+2)\left(\frac{\varphi_c}{M_P}
\right)^{-2}.
\end{eqnarray}
The spectral index calculated in the stochastic 
approach is shown in figure \ref{fig1} (b) for a range of 
values for $N_W$.
The exact value of $N_W$ depends on the history of the 
universe. Typically $N_W$ is assumed to be the number of e-folds
before the end of inflation when the present Hubble scale left the horizon,
which should be a good approximation here.
In \cite{al} it has been argued that the maximum range is given by
$14<N_W<75$. Other authors assume $N_W$ to be in the range between 
50 and 60, see for example, \cite{e-fol}, or between 46 and 60 \cite{kin}.
To see how values change in the case at hand we have chosen 
$N_W$ to be between 45 and 65.
It is found that, for a given value of $N_W$,
the larger the value $\varphi_0$ with respect to
$\varphi_W$ the smaller the spectral index.

Assuming negligible running of the spectral index and 
negligible contributions from tensor perturbations the 
spectral index is found to be \cite{wmap}
$n_s=0.951\pm0.016$ from the three year WMAP data only,
$n_s=0.948\pm 0.015$ from WMAP data and  data from the 
2dF Galaxy Redshift Survey (2df) and 
$n_s=0.948^{+0.016}_{-0.015}$ from WMAP data and data 
from the Sloan Digital Sky Survey (SDSS).
In general the spectral index in the stochastic approach 
depends on $\frac{\varphi_0}{\varphi_W}$ and $N_W$.
It is always smaller than the one obtained in the standard
approach, which only depends on $N_W$. 
This is true even at $\frac{\varphi_0}{\varphi_W}=1$.
Furthermore, it is found that only the spectral index for a potential
with $n=2$ is compatible with observations.
Moreover, in the case at hand 
the observational constraints can be satisified 
only for $N_W=55$ and $N_W=65$, requiring
$\frac{\varphi_0}{\varphi_W}<1.08$ for $N_W=55$
and 
$\frac{\varphi_0}{\varphi_W}<1.12$ for $N_W=65$.

The running of the spectral index in the standard approach \cite{ll},
$dn_{s,stand}/d\ln k=16\epsilon\eta-24\epsilon^2-2\xi_{sl}^2$, where
$\xi_{sl}^2\equiv\frac{M_P^4}{64\pi^2}\frac{V'}{V}\frac{V'''}{V}$
is a third slow roll parameter, is given by
\begin{eqnarray}
\frac{dn_{s,stand}}{d\ln k}=-\frac{n^2(n+2)}{32\pi^2}\left(
\frac{\varphi_c}{M_P}\right)^{-4}.
\end{eqnarray}
The running of the spectral index in the 
stochastic approach is shown in figure \ref{fig1}(c).
Thus as in the standard approach, the running of the spectral
index calculated in the stochastic approach is negative.
For a chosen value of $N_W$, its modulus is increasing with increasing values of
$\frac{\varphi_0}{\varphi_W}$.
\begin{figure}[ht]
\centerline{\epsfxsize=2.2in\epsfbox{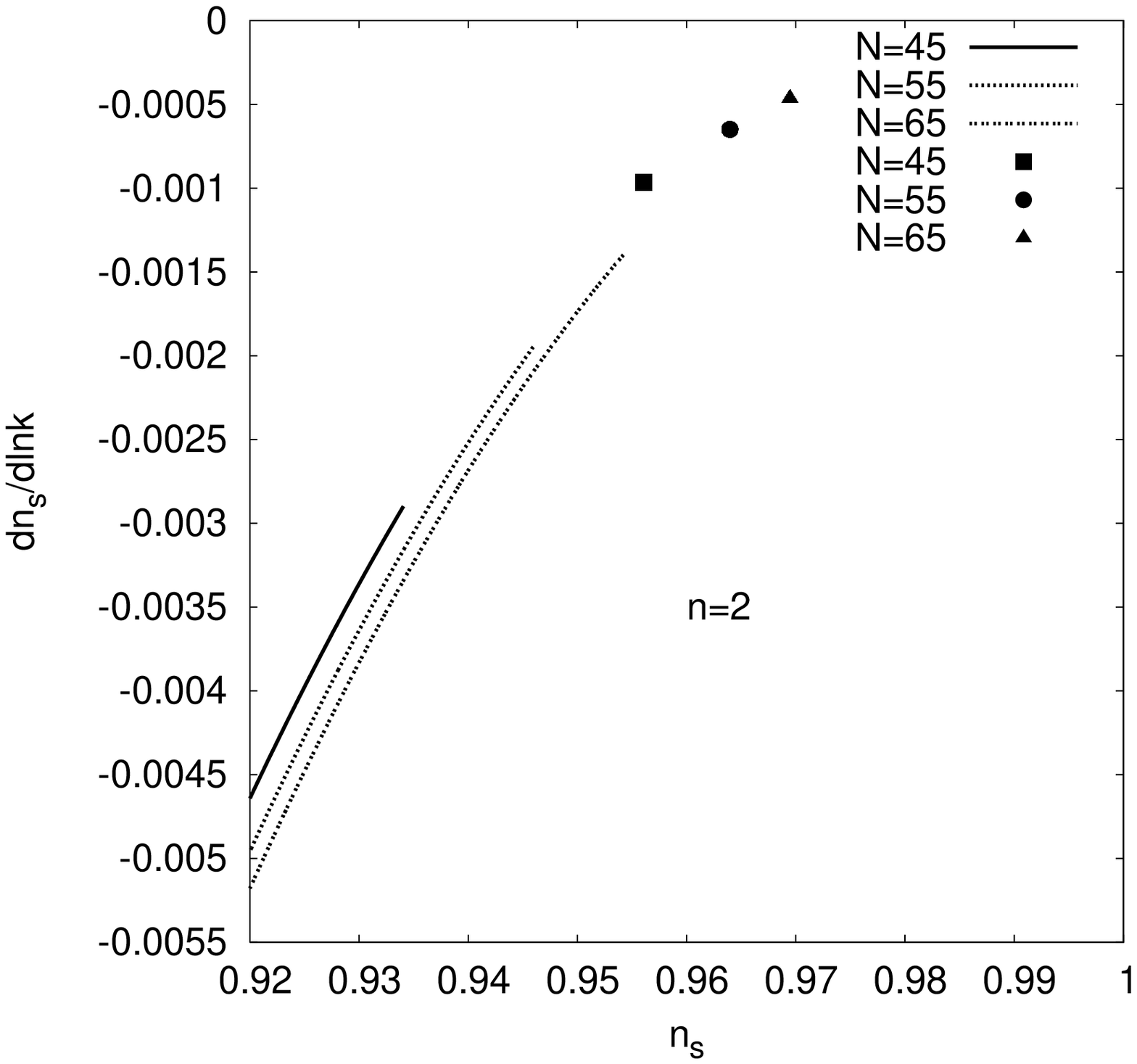}}
\caption{
The graph shows the running of the spectral index against the spectral index
in the stochastic approach (lines) and the standard approach (points)
for a quadratic potential for $N_W=45,55,65$.
} 
\label{fig1b}
\end{figure}
In figure \ref{fig1b} we have plotted in the case of a quadratic potential 
the running of the spectral index $dn_s/d\ln k$ versus the spectral
index $n_s$ in the stochastic as well as the standard approach.
As can be seen in this figure the corresponding values for the running of the 
spectral index are much below observational bounds,
assuming negligible contributions from tensor perturbations,
$\frac{dn_s}{d\ln k}=-0.055^{+0.030}_{-0.031}$ from 
WMAP three year data only, $\frac{dn_s}{d\ln k}=-0.048\pm 0.027$
from WMAP and 2df data and $\frac{dn_s}{d\ln k}=-0.060\pm 0.028$
from WMAP and SDSS data \cite{wmap}.
Thus the assumption of negligible running of the spectral 
index is well justified.
Similar results are found for the contribution from 
tensor perturbations (see next section) as can be seen from 
figure \ref{fig2} (b), which shows the tensor-to-scalar amplitude 
ratio versus the spectral index in the case of a quadratic potential
in the standard and in the stochastic approach.
Thus in order to satisfy observational bounds
$\frac{\varphi_0}{\varphi_W}$ has to be close to 1.
Therefore deviations from de Sitter space-time are small.

\section{Gravitational wave spectrum}

In this section the tensor-to-scalar amplitude ratio $r$ is 
calculated which is another observational parameter.
The gravitational wave amplitudes $h_{ij}$ are effectively
determined by two massless scalar fields $\psi_{+}$, $\psi_{\times}$,
respectively \cite{ll}.
In particular, $h_{ij}=h_{+}e_{ij}^{+}+h_{\times}e^{\times}_{ij}$, where
$e^{+}$ and $e^{\times}$ are polarization tensors and 
$h_{+,\times}=\left(\frac{M_P^2}{16\pi}\right)^{-\frac{1}{2}}
\psi_{+,\times}$. For simplicity, the index of $\psi_{+,\times}$ is omitted
and the scalar fields are simply denoted by $\psi$.
Assuming that $\psi$ is in the slow roll regime,
at first order the following equation holds 
\begin{eqnarray}
\dot{\psi}=\frac{H^{\frac{3}{2}}(\varphi_c)}{2\pi}\xi(t),
\end{eqnarray}
where the noise $\xi(t)$ is defined as before.
This is solved by 
\begin{eqnarray}
\psi(t)=\frac{1}{2\pi}\int_0^td\tau H^{\frac{3}{2}}(\tau)\xi(\tau),
\end{eqnarray}
implying
\begin{eqnarray}
\langle\psi^2(t)\rangle=\frac{1}{4\pi^2}\int_0^td\tau H^3(\tau).
\end{eqnarray}
Thus the spectrum of gravitational waves,
${\cal P}_{grav}=\frac{1}{H}\frac{d\langle h_{ij}h^{ij}\rangle}
{dt}$,
is found to be 
\begin{eqnarray}
{\cal P}_{grav}=\frac{16}{\pi}\left(\frac{H}{M_P}\right)^2,
\end{eqnarray}
which equals the spectrum in the standard approach (cf. \cite{ll}, e.g.).
Hence the tensor-to-scalar amplitude ratio $r\equiv\frac{{\cal P}_{grav}}
{{\cal P}_{\cal R}}$ is found to be 
\begin{eqnarray}
r_{stoch}=\frac{16\epsilon}{1-\frac{n}{4}\left[\left(\frac{\varphi_0}
{\varphi_c}\right)^4-1\right]}.
\end{eqnarray}
Therefore, for $\left(\frac{\varphi_0}{\varphi_c}\right)=1$
the tensor-to-scalar amplitude ratio in the standard approach, 
$r_{stand}=16\epsilon$, (see for example \cite{bkmr}) is recovered.
For all other values, $r_{stoch}>r_{stand}$.
This is due to the fact that in the stochastic approach the curvature
spectrum amplitude is lower than in the standard approach.

The tensor-to-scalar amplitude ratio calculated in the 
stochastic approach is shown
in figure \ref{fig2} (a) for different values of $N_W$.
Assuming negligible running of the spectral index 
observational bounds on $r$ are found to be:
$r<0.65$ from the three year WMAP data only, $r<0.38$ from WMAP and 2df 
data and $r<0.30$ from WMAP and SDSS data (\cite{wmap} (for the last
value see also \cite{kin}).
The constraint from the WMAP three year data only admits 
all three types of potentials
considered in figure \ref{fig2} (a), up to a certain value of 
$\frac{\varphi_0}{\varphi_W}$. The constraint from WMAP and 2df data marginally 
allows a quartic potential with $N_W=45$, which is already excluded by 
the WMAP plus SDSS data, independent of $\frac{\varphi_0}{\varphi_W}$.
This stronger observational constraint can be satisfied up to
a certain value of $\frac{\varphi_0}{\varphi_W}$ for potentials
with $n=4$ and $N_W>45$ and for quadratic potentials.
In figure \ref{fig2} (b) the tensor-to-scalar amplitude ratio versus
the spectral index is shown for a quadratic potential
calculated in the stochastic as well as the standard approach. 
Values for $r$ allowed by the observational constraints on the spectral index are 
of the same order in the stochastic and in the standard approach.

\section{Non-gaussianity}
By construction, the first order perturbation $\delta\varphi_1$ is gaussian
with zero mean, $\langle\delta\varphi_1\rangle=0$.
Thus the curvature perturbation ${\cal R}$ is gaussian at first order.
Non-gaussianity enters in ${\cal R}$ at second order in the 
expansion of $\delta\varphi$ (cf. equation (\ref{e2})).
The nonlinearity can be estimated using the parameter 
$f_{NL}$ defined by \cite{fnl}
\begin{eqnarray}
{\cal R}={\cal R}_{g}-\frac{3}{5}f_{NL}\left(
{\cal R}_g^2-\langle{\cal R}_g^2\rangle\right),
\end{eqnarray}
where ${\cal R}_g$ is a gaussian distribution with 
$\langle {\cal R}_g\rangle=0$. Thus by construction 
$\langle {\cal R}\rangle=0$.
In order to determine the non-gaussian contribution
due to the second order perturbation, it is 
convenient to consider the curvature perturbation
${\cal R}=-\frac{H}{\dot{\varphi}_c}\delta\varphi$,
where $\delta\varphi=\delta\varphi_1+\delta\varphi_2-\langle\delta\varphi_2
\rangle$. Hence by construction $\langle{\cal R}\rangle=0$.
Furthermore, the gaussian contribution is given by
\begin{eqnarray}
{\cal R}_g\equiv-\frac{H}{\dot{\varphi}_c}\delta\varphi_1.
\end{eqnarray}
Thus $f_{NL}^2$ is found to be  
\begin{eqnarray}
f_{NL}^2=\frac{25}{9}\left(\frac{M_P^2}{4\pi}\right)^2
\left(\frac{H'}{H}\right)^2
\frac{\langle\delta\varphi_2^2\rangle-\langle\delta\varphi_2\rangle^2}
{\langle\delta\varphi_1^4\rangle-\langle\delta\varphi_1^2\rangle^2}.
\end{eqnarray}
Taking into account only the dominant contribution in $\delta\varphi_2$,
coming from the second term in equation (\ref{p2}),
yields to 
\begin{eqnarray}
f_{NL}^2\simeq\frac{25}{128\pi^2}\left(\frac{H}{M_P}\right)^{-2}
\frac{\int_{\varphi_c}^{\varphi_0}d\psi HH'\int_{\psi}^{\varphi_0}
d\chi\left(\frac{H}{H'}\right)^3
+\frac{M_P}{4\pi}\int^{\varphi_0}_{\varphi_c}d\psi\frac{H^4}{H'}}
{M_P^{-2}\left(\int_{\varphi_c}^{\varphi_0}d\psi\left(\frac{H}{H'}
\right)^3\right)^2}.
\label{fn1}
\end{eqnarray}
This expression is exact for potentials with index $n=2$ and 
$n=4$, since in these cases $H'''\equiv 0$.
For a potential of the form $V=V_0\left(\frac{\varphi_c}{M_P}\right)^n$ 
equation (\ref{fn1}) leads
to, at $\varphi_c=\varphi_W$,
\begin{eqnarray}
f_{NL}^2&\simeq&
\frac{25}{128\pi^2}
\left[
\frac{n^4}{4}\left[
\frac{4}{n\left(n+4\right)}\left(\frac{\varphi_0}{\varphi_W}
\right)^{n+4}-\frac{1}{n}\left(\frac{\varphi_0}{\varphi_W}\right)^4
+\frac{1}{n+4}\right]\left(\frac{\varphi_W}{M_P}\right)^{-4}
\right.
\nonumber\\
&&
\left.
+
\frac{n^5}{4\pi(3n+4)}\left(\frac{8\pi}{3}\right)^{\frac{1}{2}}
\left(\frac{V_0}{M_P^4}\right)^{\frac{1}{2}}
\left(\frac{\varphi_W}{M_P}\right)^{\frac{n}{2}-6}
\left[\left(\frac{\varphi_0}{\varphi_W}\right)^{\frac{3n}{2}+2}
-1\right]
\right]
\left[\left(\frac{\varphi_0}{\varphi_W}\right)^4-1\right]^{-2}
\end{eqnarray}

\begin{figure}[ht]
\centerline{\epsfxsize=2.2in\epsfbox{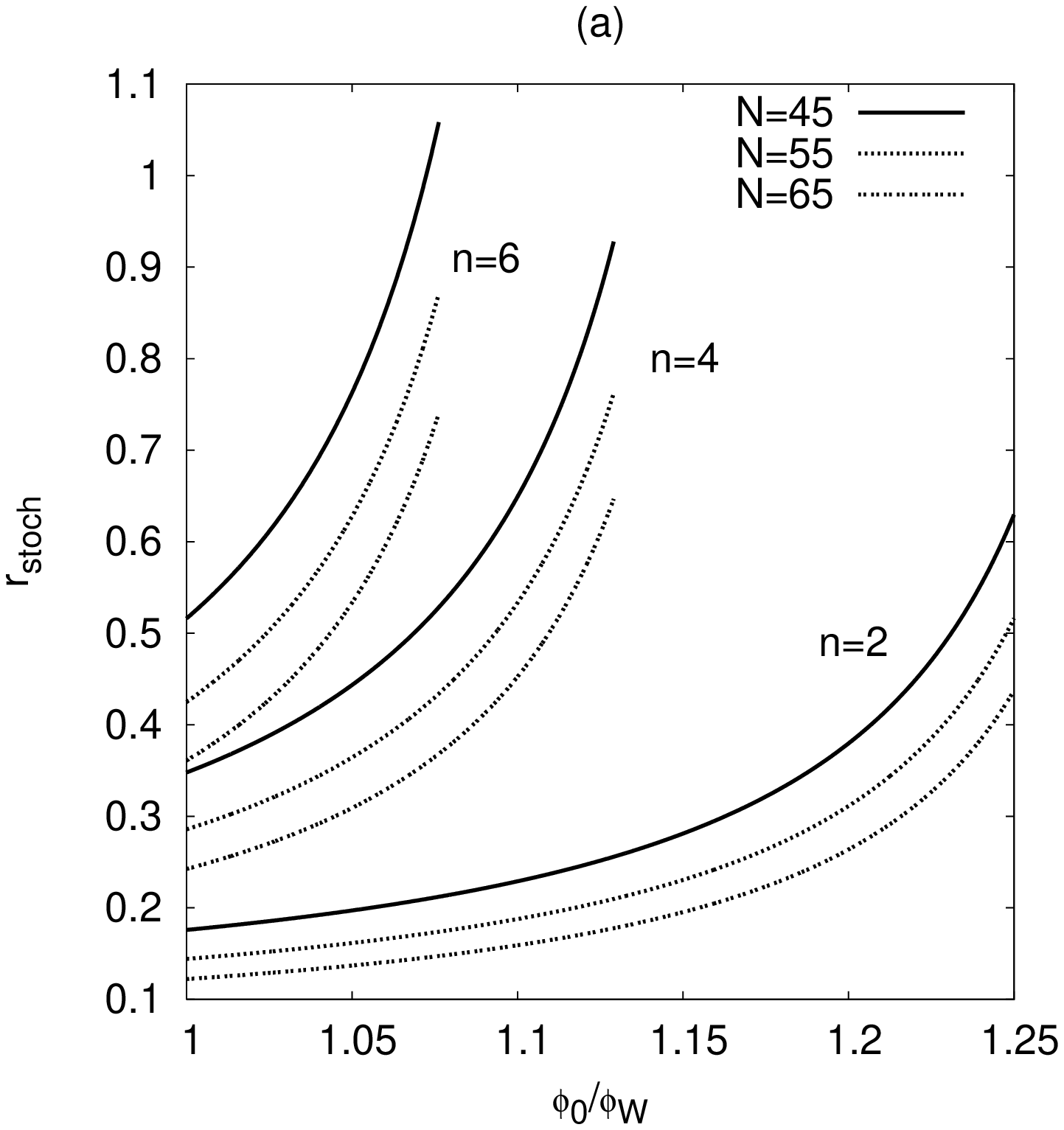}
\hspace{0.05cm}
\epsfxsize=2.2in\epsfbox{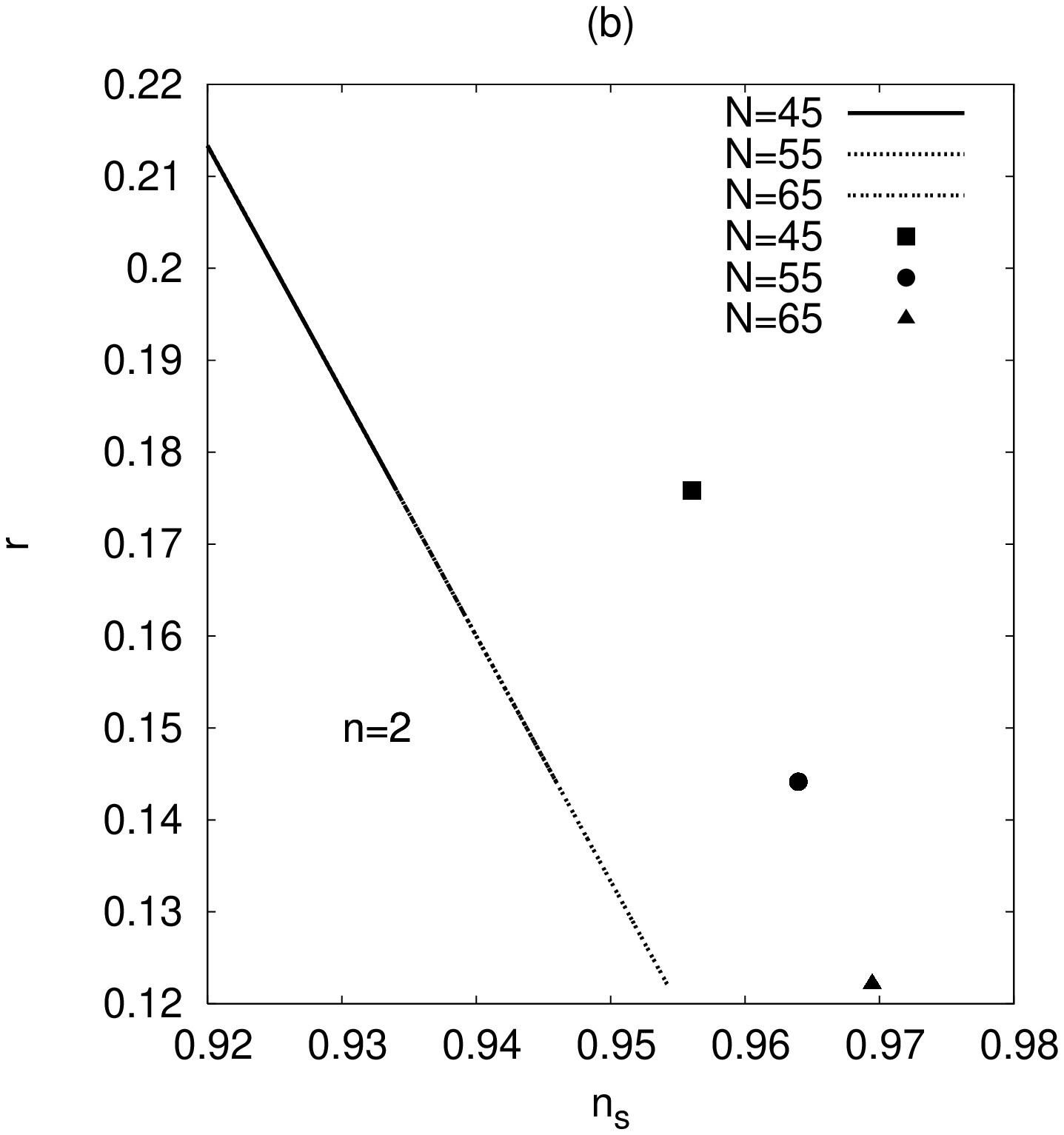}
\hspace{0.05cm}
\epsfxsize=2.2in\epsfbox{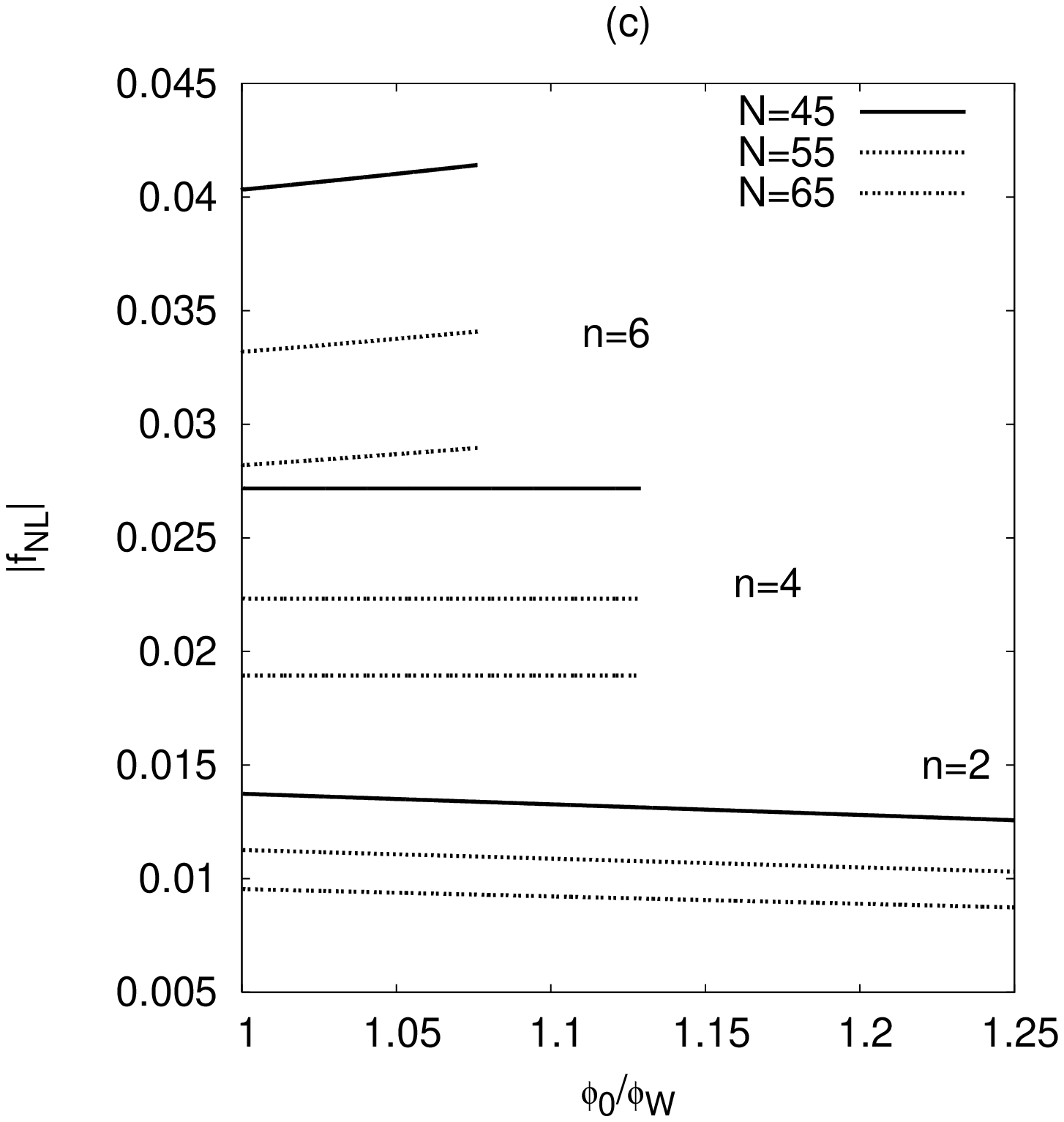}
}
\caption{(a) The tensor-to-scalar amplitude ratio $r_{stoch}$ in  the 
stochastic approach.
(b) The tensor-to-scalar amplitude ratio 
against the spectral index in the stochastic (lines)
and in the standard approach (points) in the case of a 
quadratic potential.
(c) The non-gaussianity parameter $|f_{NL}|$ in the stochastic
approach.
Curves are shown for $N_W=45,55,65$. Furthermore, $n$ denotes the 
power of the potential. 
}
\label{fig2}
\end{figure}

The parameter $|f_{NL}|$ is shown in figure \ref{fig2} (c)
for several values of $N_W$ and the potential index $n$. 
An estimate for $\left(\frac{V_0}{M_P^4}\right)$ is found using 
the expression for the power spectrum amplitude 
${\cal P}_{\cal R}$ (cf. equation (\ref{n1})) together with 
${\cal P}_{\cal R}=2.4\times 10^{-9}$ \cite{wmap}.
As can be seen from figure \ref{fig2} (c),
$f_{NL}$ satisfies the bound from observations, 
$-54<f_{NL}<114$ \cite{sper}.

For $n=4$ the parameter $|f_{NL}|$ only depends very weakly 
on $\frac{\varphi_0}{\varphi_W}$. For potentials with an index $n>4$ 
the parameter $|f_{NL}|$ is increasing 
with increasing values of $\frac{\varphi_0}{\varphi_W}$.
For $n=2$ $|f_{NL}|$ is a descreasing function 
for increasing values of $\frac{\varphi_0}{\varphi_W}$.
The non-gaussian contribution requires that $\frac{\varphi_0}{\varphi_W}>1$.

These results can be compared with values for $f_{NL}$ found in the standard
approach. Perturbation analysis up to second order in the standard 
approach to slow roll inflation leads to the estimate that $f_{NL}$ is
of the order of the slow roll parameters \cite{mald}-\cite{sl}.
In \cite{mald} $f_{NL}$ is found to be $f_{NL}\sim\frac{5}{12}(n_s+f(k)n_t)$,
where $n_t$ is the spectral index of the gravitational wave spectrum and 
$0\leq f\leq\frac{5}{6}$ characterizes the configuration of the momenta which 
enters into the calculation of the three-point correlation function of the 
curvature perturbation.
Choosing $f_{NL}=\frac{5}{6}$ (equilateral triangle) results for 
the range of parameters under consideration here ($N_W=45,55,65$ and $n=2,4,6$)
in a non-linearity parameter $f_{NL}\sim {\cal O}(10^{-2})$. Thus it is 
of the same order of magnitude as $f_{NL}$ calculated in the
stochastic approach (cf.  figure \ref{fig2} (c)).

\section{Conclusions}

Recently, the equations of stochastic inflation have 
been solved in a perturbative way \cite{sto-pert}.
This allows to take into account the effect
on the perturbations due to the evolving 
Hubble parameter.
Here, the perturbative approach to stochastic inflation 
has been used to determine the spectrum of the
curvature perturbation due to the 
coarse grained inflaton as well as the 
gravitational wave spectrum.

The spectrum of curvature fluctations has been calculated 
using the solution for the perturbation of the 
coarse grained field up to first order in the noise.
The resulting expression for the spectrum contains 
an additional parameter in the form of a value $\varphi_0$
of the coarse grained scalar field $\varphi$. This has been 
interpreted as the value of the deterministic 
coarse grained inflaton $\varphi_c$ at the horizon.
This interpretation is motivated by the fact that the 
resulting perturbation spectrum reduces to the 
standard expression in the case  $\varphi_c=\varphi_0$.
In the standard approach the curvature perturbation 
spectrum is found using the spectrum of vacuum fluctations
in a de Sitter background calculated at the horizon.
The coarse graining of the scalar field in the stochastic
approach to inflation introduces a coarse graining scale.
This is at least of the order of the horizon.
The spectrum is calculated at the coarse graining scale.
In general the amplitude of the spectrum of curvature 
fluctuations calculated using the stochastic approach 
is reduced, in comparison with the standard approach,
for $\frac{\varphi_0}{\varphi_c}>1$.
This results from the 
horizon and the coarse graining scale  not being the same
in this case.
Modes with wavelengths between the horizon and 
the coarse graining scale do not contribute to the 
spectrum.

The expression for the spectral index depends in general on 
the parameter $\varphi_0$. Even at $\varphi_c=\varphi_0$ the
spectral index is smaller than the one obtained in the
standard approach. Moreover, only for polynomial potentials
with a potential index $n=2$ observational bounds can be satisfied.
The running of the spectral index calculated in the stochastic 
approach is negative as it is the case in the standard approach.
Apart from its dependence on the index of the potential and 
the number of e-folds before the end of inflation, at which the 
coarse grained field is evaluated, its value also depends 
on $\frac{\varphi_0}{\varphi_c}$.

The tensor-to-scalar amplitude ratio is in general 
larger in the stochastic than in the standard approach
for $\frac{\varphi_0}{\varphi_c}>1$.
This is due to the fact that whereas the gravitational wave amplitude
is found to be the same as in the standard approach 
the scalar perturbation amplitude is lower in the 
stochastic approach.

Finally, the non-gaussian contribution due to the second order
perturbation in the stochastic approach to inflation 
has been calculated. By construction perturbations at
first order in the noise are gaussian since the 
noise is gaussian. Non-gaussianity enters at second order
in the noise. The resulting expression for the parameter 
$|f_{NL}|$ reqires $\frac{\varphi_0}{\varphi_c}>1$.
For potentials with a potential index $n=2$ it is 
a decreasing function with increasing 
$\frac{\varphi_0}{\varphi_c}$, for $n=4$ it is only very weakly
depending on the value of $\frac{\varphi_0}{\varphi_c}$
and for $n>4$ it is an increasing function with  $\frac{\varphi_0}{\varphi_c}$.
The resulting values for $|f_{NL}|$ are compatible with bounds from 
observations.

\section{Acknowledgements}
 
I would like to thank H.-P. Breuer, D. Wands and in particular
M. A. V\'azquez-Mozo for
useful discussions. 
Furthermore, I am grateful to N. Afshordi for 
valuable comments on
an earlier version of this paper.
I am grateful to the theory 
division at CERN for hospitality where part of this 
work was done.
This work has been supported by the programme
``Ram\'on y Cajal'' of the M.E.C. (Spain).
Partial support by Spanish Science Ministry
Grants FPA2005-04823
and BFM 2003-02121 is acknowledged.

\end{document}